# Enhancement of interferometric precision using fast light

*M.S. Shahriar, R. Tripathi, G.S. Pati, V. Gopal, M. Messall and K. Salit*

*ECE Department, Northwestern University*

*2145 N. Sheridan Rd, Evanston IL 60208*

We show that anomalous dispersion characteristic of fast-light can be used to enhance the sensitivity of optical interferometry under certain conditions. In particular, we show that a dual-chamber Fabry-Perot interferometer with a shared mirror-pair can be used in a way so that its sensitivity is increased by operating near the critically anomalous dispersion condition where the group index is much less than unity. The enhancement factor can be as high as $10^8$ for realistic conditions. The process of bi-frequency pumped Raman gain in a $\Lambda$-type atomic medium can be used to achieve this effect.

Metrological applications deal with ultra-precision standards that require measurements using optical interferometers with extremely high sensitivity [1-4]. Recent experiments have shown that 'slow' or 'fast' group velocities in different kinds of material medium can drastically enhance the dispersion property under resonance condition [5-9]. It has been predicted that nonlinear Kerr index change of an active material in a photonic crystal based waveguide Mach-Zehnder interferometer (MZI) system can be dramatically enhanced using slow group velocity of light [10].

Increased nonlinear phase sensitivity can potentially enable designs for extremely small size integrated optic switches, routers and wavelength converters etc. [11,12]

However, it has been generally perceived that the slow group velocity in a highly linear dispersive medium does not play any role in an optical interferometer so far as precision interferometric sensing or measurements are concerned. This conclusion holds, for example, for the MZI shown in figure 1. The basic argument stems from the fact that the group velocity represents the speed of the envelope of a pulse, while the optical carrier wave underneath propagates with the phase velocity, and the interference process is insensitive to the behavior of the envelope. This conclusion also holds in rotation sensing using a Sagnac interferometer where the medium is co-rotating with the interferometer. However, it has been shown recently [13,14] that if the medium is not co-rotating with the interferometer, for a device that can be used for relative rotation sensing, the use of slow light leads to a strong enhancement in the rotational sensitivity. Furthermore, we have shown [14] that in a passive optical ring resonator/interferometer, commonly used in optical gyroscopes, fast or 'superluminal' light propagation near critically anomalous dispersion (CAD) can enhance the sensitivity of measuring absolute rotation by many orders of magnitude. In this paper, we show that a similar fast light induced enhancement can also be achieved in non-Sagnac interferometry.

Specifically, we consider a dual-chamber Fabry-Perot (FP) resonator with the optical beams sharing a common mirror-pair configuration, with both filled with a fast-light medium but only one of the chambers being subject to the perturbation to be measured. The strong index dispersion characteristic of superluminal propagation can produce enhanced frequency shifts proportional to extremely small index variation in the test chamber. The large, spectrally narrow negative dispersion slope in the vicinity of the CAD condition can give rise to sensitivity enhancement factor as high as $10^8$. The characteristic dispersion necessary is experimentally obtainable using the gain transparency induced by bi-frequency pumped Raman gain doublets in a Λ-type atomic medium. We discuss such a scheme in detail, and also predict a maximum sensitivity for the interferometer using such a process by estimating the magnitude of the second-order dispersion. We also believe that the present idea can be extended to construct fiber-based precision interferometric sensor, using similar dispersive phenomena observed under stimulated Brillouin loss [15] during light propagation in fibers.

The basic configuration of this CAD-enhanced Fabry-Perot (FP) Interferometer is illustrated in figure 2. The segmented volume inside the resonator consists of two parts. The part on the right constitutes the reference volume and the one on the left is the test chamber which is exposed to physical processes that cause the index change to be sensed. The reference volume is shielded in a way so that it does not see the effect of the physical parameter S. In particular, consider a test medium for which the index changes linearly as a function of S, independent of frequency over a small frequency bandwidth. This can happen, for example, in an alkali atom vapor when the temperature of the cell is changed. In this case, the temperature plays the role of the parameter S. The frequencies of the two input optical fields can be tuned independently using acousto-optic modulators (AOMs), as shown. A null-point can be established independently at the peak of the FP resonance in each chamber, using independent feedback mechanisms. If the optical fields see identical effects in the medium in each region, the transmitted outputs can be tuned by the feedback mechanism to the peak of FP resonance using degenerate frequency fields. However, when the perturbation parameter S is non-zero, two distinct frequencies are transmitted through these regions when the transmitted outputs are again tuned to the peaks of the FP resonances. Under this condition, the beat note of these frequencies Δf which is a direct measure of the physical process can be measured from heterodyne detection at the output of the interferometer. This model can be represented quantitatively by expressing the refractive indices in these regions as follows:

$$\text{ref region}: n(\omega) = n_o + \Delta\omega \frac{\partial n}{\partial \omega}$$

$$\text{test region}: n(\omega) = \left(n_o + \Delta S \frac{\partial n}{\partial S}\right) + \Delta\omega \frac{\partial n}{\partial \omega}; \quad \{\partial n / \partial S \equiv \sigma, \text{independent of } \omega\} \quad (1)$$

$$= n' + \Delta\omega \frac{\partial n}{\partial \omega}$$

where $n_o$ is the mean refractive index of medium that is considered the same for both. In case of dilute atomic medium under resonant excitation, $n_o$ is close to unity. If we ignore the effect of dispersion, the resonant frequency for each zone are respectively given by

$$\omega_o = C_o/(2n_o L), \quad \omega_o' = C_o/(2n' L) = \omega_o \left(1 - \frac{\sigma}{n_o}\Delta S\right), \quad \left|\frac{\Delta S}{n_o}\frac{\partial n}{\partial s}\right| \ll 1 \quad (2)$$

where L is the distance between the two resonator mirrors. When S is non-zero, if we consider the case when there is no dispersion, the beat frequency is given by

$$\Delta\omega = \omega_o \frac{\sigma}{n_o}\Delta S \equiv \Delta\omega_o \quad (3)$$

It also confirms the fact that when S=0, the two chambers are identical so that the beat frequency is zero, as it should be. The expression for the beat frequency in eqn. 3 is strictly derived assuming phase index to be independent of frequency i.e. when the medium exhibits no dispersion. As we will show next, when the effect of dispersion is taken into account, the result changes significantly. In what follows, we derive this result and discuss the strong implications thereof.

While calculating the velocities of the optical phase fronts ($c_o/n$), we expand the index n around ω and also assume the magnitude of dispersion $\left|\frac{\partial n}{\partial \omega}\right|$ in the medium contained in the reference chamber is extremely large compared to that in the test chamber. Under this condition, one can obtain a self-consistent relation for beat frequency $\Delta\omega$ as follows

$$\Delta\omega = \frac{\Delta\omega_o}{1+\tilde{n}} = \Delta\omega_o\, \xi, \quad \tilde{n} \equiv \frac{1}{n_o}\omega_o\left(\partial n/\partial\omega\right), \quad \xi = \frac{n_o}{n_g}, \quad n_g = n_o + \omega_o\left[\frac{\partial n}{\partial \omega}\right] \qquad (4)$$

For a medium that produces the so-called fast light under strong anomalous dispersion condition ($\partial n/\partial\omega < 0$), can give rise to $n_g \ll n_o$ [i.e., $\partial n/\partial\omega \ll (n_o/\omega_o)$], so that this result implies an enhancement in interferometric sensitivity. It is easily possible to achieve a condition where $0 < n_g \ll 1$ (or $\xi \gg 1$) that implies a potentially significant enhancement of sensitivity. Note that this enhancement happens nears the vicinity of the dispersion condition such that $\partial n/\partial\omega = -(n_o/\omega_o)$, the physical process we designate as critically anomalous dispersion (CAD) condition. Thus, the sensitivity of the sensor can be enhanced by a very large factor as long as the value of $n_g$ is near the vicinity of the CAD condition. Experiments have been performed recently using large spectrally narrow negative dispersion slope of this type to demonstrate superluminal light propagation, without violating causality and special relativity the causality [6,7]. As such, the CAD condition necessary for implementing this interferometer can be readily obtained experimentally.

The enhancement factor ξ becomes unphysical when the group index $n_g$ approaches zero. In an experimental situation, this does not happen since the second or higher order dispersions which are ignored in our analysis sets the upper bound for enhancement. In order to quantify it, we extend our previous analysis to include the second order dispersion and obtain the expression for the beat frequency Δω, given by

$$\Delta\omega = \Delta\omega_o \cdot \eta = \Delta\omega_o \cdot \left| \frac{\pm 2\xi}{1 \pm \sqrt{1+Q\xi^2}} \right|, \quad Q = f\, n''\, \omega_o^2, \quad n'' \equiv [\partial^2 n / \partial\omega^2]/n_o \qquad (5)$$

where $f = \dfrac{\Delta\omega_o}{\omega_o}$ is a dimensionless quantity defined as the fractional bandwidth. In the limit when $\xi \to \infty$ or ($n_g \to 0$), non-zero value of Q (assuming $|Q| \ll 1$) governed by magnitude of second order dispersion prevents the enhancement factor near the CAD condition from becoming unphysical. Figure 3 shows a plot for sensitivity enhancement over a restricted range of (1/ξ = $n_g/n_o$) values choosing appropriate signs from in eqn. (5) for positive and negative values of $n_g$. The enhancement bound saturates to a maximum value of $\eta_{mac} = 2/\sqrt{|Q|}$.

We point out here that the process of CAD induced enhancement summarized here is in many ways similar to the corresponding enhancement we have predicted for a passive resonator based optical gyroscope that makes use of the Sagnac effect [14]. In the case of the later, the fact that both the clock-wise and the counter-clock-wise modes in the resonator share a common cavity is the critical factor. Similarly, in the case of the former, the crucial factor that makes the enhancement possible is the fact that both chambers share the same mirror pair.

In what follows, we briefly recall an experimental scheme for producing large anomalous dispersion near gain transparency condition using bi-frequency pumped Raman gain doublet (BPRGD) in a Λ-type atomic vapor, for example. Figure 4a shows a typical Λ system consisting of two metastable ground states (|1> and |3>) coupled to an excited state (|2>) through electric dipole interactions. The pump frequency is also detuned from atomic transition frequency by more than the Doppler width so that atoms mostly remain in the ground states. In the presence of an optical pumping beam (not shown), steady-state Raman-type population inversion occurs between states |1> and |3>) that induces gain in a probe field when the two-photon resonance condition is satisfied [6, 16] between the single pump and the probe. The gain at the probe frequency is due to the Raman transition between |1> and |3> that results in the absorption of a pump photon followed by emission of a photon at the probe frequency. We note parenthetically that the initial population inversion between the ground states can be produced by self-optical pumping due to the Raman pump instead of an external optical pump [16].

If there are two close frequencies present in the pump, they can produce two narrow gain peaks, known as Raman gain doublets, each corresponding to the two photon resonance condition for one of the pumps [6]. From the Kramer-Kronig relations, it then follows that the index profile displays anomalous dispersion (i.e., $\partial n/\partial \omega < 0$) at the center of these two gain peaks. Figure 4b shows the frequency dependence of gain coefficient and anomalous dispersion under gain transparency near the vicinity of zero probe detuning that corresponds to dn/dω = -3.1x10$^{-16}$ rad$^{-1}$sec, d$^2$n/dω$^2$ = 4.1x10$^{-38}$ rad$^{-2}$sec$^2$, and $\eta_{max}$=8.0x10$^7$. The slope of the dispersion can be tuned by controlling the strength or the frequency separation of the pumps, thus enabling one to tune the dispersion near the CAD condition [ $\partial n/\partial \omega = -(n_o/\omega_o)$ ] in order to achieve very large sensitivity enhancement. Assuming a Doppler-free situation for simplicity (valid for pump detuning much larger than the Doppler width), the magnitudes of the gain and the associated dispersion can be estimated by using a perturbative expression for the optical susceptibility [6]:

$$\chi(\omega) = \frac{N|d_{32}|^2}{4\pi\hbar\varepsilon_o \Delta_o^2}\left[\frac{\Omega_1^2}{\omega-\omega_1+i\Gamma} + \frac{\Omega_2^2}{\omega-(\omega_1-\Delta)+i\Gamma}\right] \tag{6}$$

where $\Omega_{1,2}$ are the Rabi frequencies of the Raman pump fields, $\Delta$ and $\Delta_o$ are the average and difference detuning between the pump fields, $d_{32}$ is the dipole matrix element for the $|3>$-$|2>$ transition, $\Gamma$ is the inverse of the Raman transition lifetime and N is the atomic density. Figure 4b also shows a typical gain-assisted anomalous dispersion profile that corresponds to a very small group index $n_g < 1$ at the center of two gain profiles. This is obtained by estimating the magnitude of linear anomalous dispersion ($\partial n/\partial \omega \sim -3.1 \times 10^{-16}$ rad$^{-1}$.sec) at the center of the gain peaks. The bounded sensitivity enhancement factor, $\eta_{max}$, for the interferometer is estimated from the magnitude of the second order dispersion over a frequency bandwidth of 1 MHz, for example, and is found to be $\sim 10^8$. Note that the value of $\eta_{max}$ actually increases for smaller bandwidths, since the role of the second order dispersion gets diminished.

In an experiment, the closed-loop performance of the interferometer will be governed by the signal-to-noise (SNR) ratio in the measurement of the heterodyne beat note $\Delta f$. The uncertainty in measurement of $\Delta f$ will limit the sensitivity of the interferometer. For an ideal, shot-noise limited detection, the uncertainty in sensing is given by

$$\delta S = 2\pi(\delta f)\frac{n_o}{\sigma} = \frac{2\pi n_o}{\sigma}\left[\frac{\Delta f_c}{(N_{ph}\eta_D \tau)^{\frac{1}{2}}}\right] \tag{7}$$

where $N_{ph}$ is the average number of photons per sec incident on the detector, $\eta_D$ is the quantum efficiency of the detector, $\Delta f_c$ is the line width of cavity resonance, and $\tau$ is the integration time. The SNR in the detection process can be increased by taking measurements over long integration time if the system does not drift appreciably over the integration time. While utilizing the maximum achievable sensitivity near the CAD condition in a closed-loop feedback controlled resonator geometry, the linewidths of the voltage-controlled oscillators that drive the AOMs, as well as the and AOM beam pointing instabilities will further limit the system performance. Although here we envisaged a proof of principle experiment using a shared-mirror, common cavity configuration where the density of the test medium was varied to change the mean index, alternative geometries using stimulated Brillouin loss for fast light in fibers [15] can be constructed to design more versatile and potentially compact interferomteric systems for precision sensing and interferometry.

In conclusion, we have discussed a general purpose interferometric sensing scheme using spectrally narrow anomalous dispersion characteristic of superluminal light propagation in a specially designed FP resonator configuration. Such a scheme can potentially enhance the interferometric sensitivity by several orders of magnitude compared to a conventional interferometer and can be experimentally implemented using the near-resonant transparency condition produced by bi-frequency pumped Raman gain in an atomic medium.

This work was supported by AFOSR Grant # FA9550-04-1-0189.

**Figure Captions:**

**Figure 1:** The basic setup of a symmetric waveguide Mach Zehnder Interometer

**Figure 2:** Schematic illustration of a general purpose sensing interferometric resonator

**Figure 3:** Variation of sensitivity enhancement factor in the vicinity of CAD condition

**Figure 4: a.** Schematic illustration of the bi-frequency pumped Raman gain doublet process in a Λ-type system;

    **b.** Frequency dependence of gain coefficient and anomalous dispersion under gain transparency near the vicinity of zero probe detuning.

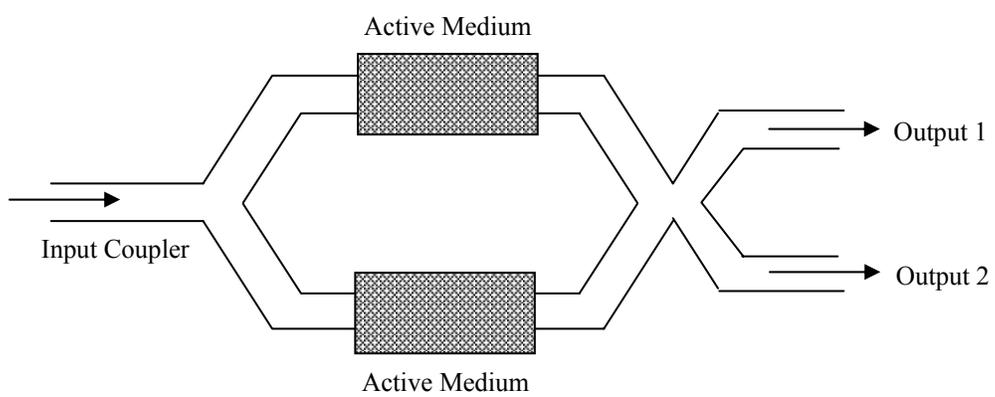

**Figure 1**: The basic setup of a Symmetric Mach Zehnder Interometer

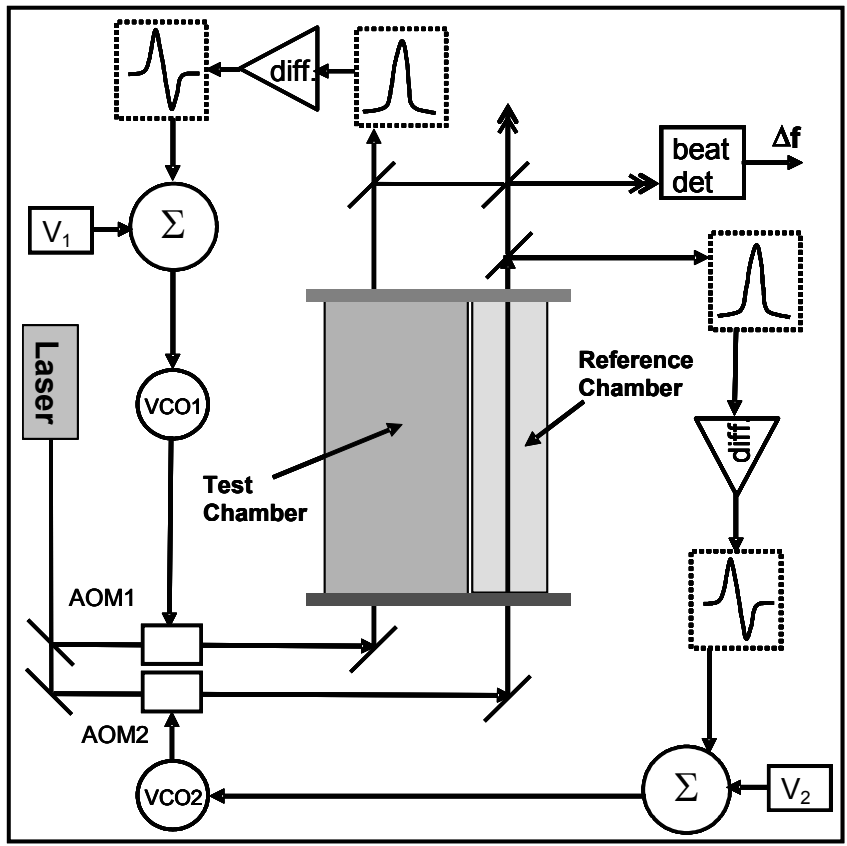

**Figure 2**: Schematic illustration of a general purpose sensing interferometric resonator

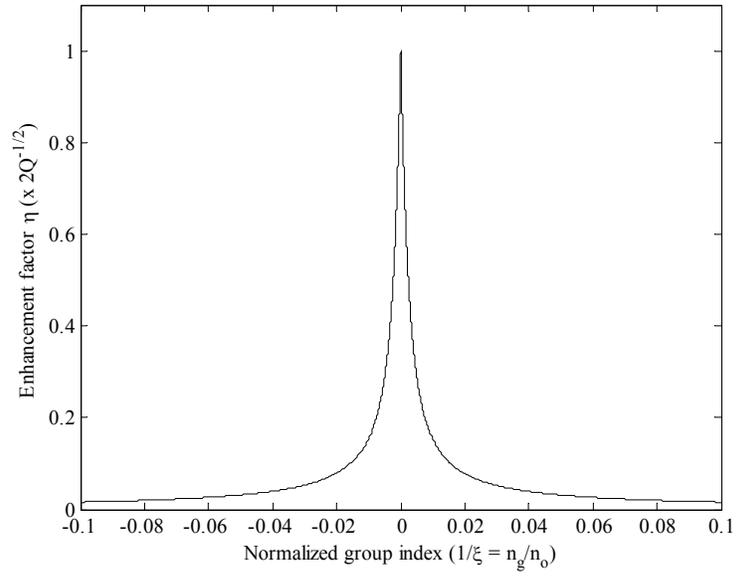

**Figure 3** Variation of sensitivity enhancement factor in the vicinity of CAD condition

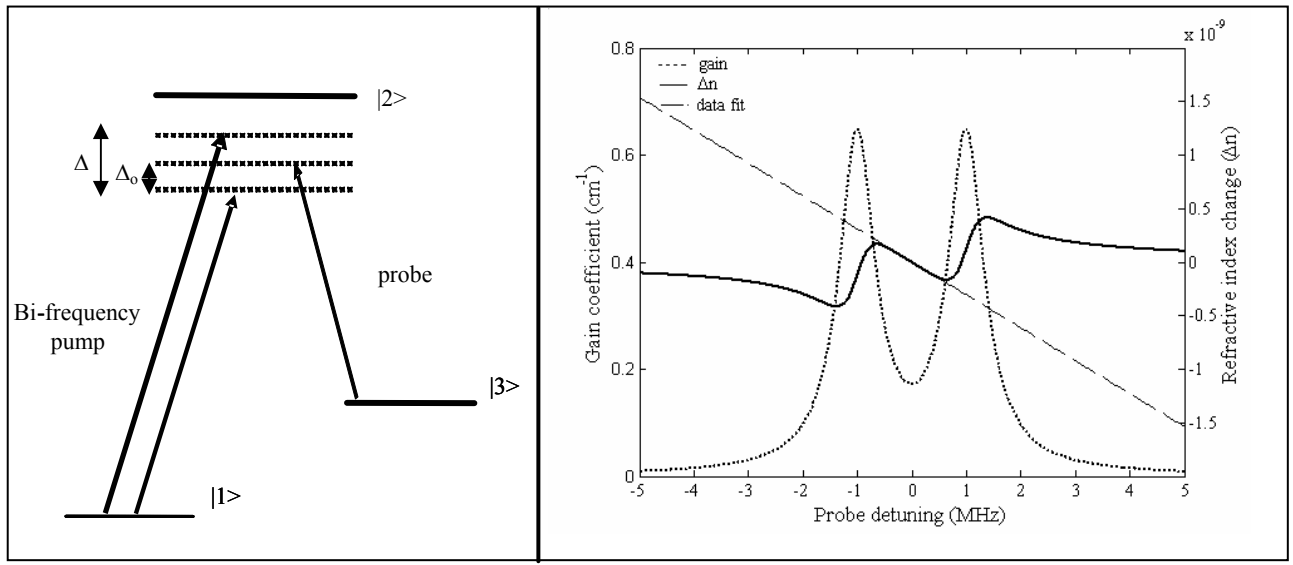

**Figure 4: a.** Schematic illustration of the bi-frequency pumped Raman gain doublet process in a Λ-type system;

**b.** Frequency dependence of gain coefficient and anomalous dispersion under gain transparency near the vicinity of zero probe detuning.